\documentclass[manuscript]{aastex}
\usepackage[utf8]{inputenc}
\usepackage{ulem}

\shorttitle{Distance to the NGC 7793 based on Cepheids}
\shortauthors{Zgirski et al.}

\begin{document}

\title{The Araucaria Project. The Distance to the Sculptor Group Galaxy NGC 7793 from Near-Infrared Photometry of Cepheid Variables}

\author{Bartlomiej Zgirski}
\affil{Nicolaus Copernicus Astronomical Center, Polish Academy of Sciences, Bartycka 18, 00-716 Warsaw, Poland}
\email{bzgirski@camk.edu.pl}
\author{Wolfgang Gieren}
\affil{Universidad de Concepcion, Departamento de Astronomia, Casilla 160-C, Concepcion, Chile
\\Millenium Institute of Astrophysics, Santiago, Chile}
\email{wgieren@astro-udec.cl}
\author{Grzegorz Pietrzynski}
\affil{Nicolaus Copernicus Astronomical Center, Polish Academy of Sciences, Bartycka 18, 00-716 Warsaw, Poland\\
Universidad de Concepcion, Departamento de Astronomia, Casilla 160-C, Concepcion, Chile}
 \email{pietrzyn@camk.edu.pl}
\author{Paulina Karczmarek}
\affil{Warsaw University Observatory, Al. Ujazdowskie 4, 00-478, Warsaw, Poland}
\email{pkarczmarek@astrouw.edu.pl}
\author{Marek Gorski}
\affil{Universidad de Concepcion, Departamento de Astronomia, Casilla 160-C, Concepcion, Chile
\\Millenium Institute of Astrophysics, Santiago, Chile}
\email{mgorski@astrouw.edu.pl}
\author{Piotr Wielgorski}
\affil{Nicolaus Copernicus Astronomical Center, Polish Academy of Sciences, Bartycka 18, 00-716 Warsaw, Poland}
\email{pwielgor@camk.edu.pl}
\author{Weronika Narloch}
\affil{Nicolaus Copernicus Astronomical Center, Polish Academy of Sciences, Bartycka 18, 00-716 Warsaw, Poland}
\email{wnarloch@camk.edu.pl}
\author{Dariusz Graczyk}
\affil{Universidad de Concepcion, Departamento de Astronomia, Casilla 160-C, Concepcion, Chile
\\ Nicolaus Copernicus Astronomical Center, Polish Academy of Sciences, Bartycka 18, 00-716 Warsaw, Poland}
\email{darek@astro-udec.cl}
\author{Rolf-Peter Kudritzki}
\affil{Institute for Astronomy, University of Hawaii at Manoa, 2680 Woodlawn Drive, Honolulu HI 96822, USA}
\authoremail{kud@ifa.hawaii.edu}
\author{Fabio Bresolin}
\affil{Institute for Astronomy, University of Hawaii at Manoa, 2680 Woodlawn  Drive, Honolulu HI 96822, USA}
\email{bresolin@ifa.hawaii.edu}

\begin{abstract}
Following the earlier discovery of classical Cepheid variables in the Sculptor Group spiral galaxy NGC 7793 from an optical wide-field 
imaging survey, we have performed deep near-infrared $J$- and $K$-band follow-up photometry of a subsample of these Cepheids to derive 
the distance to this galaxy with a higher accuracy than what was possible from optical photometry alone, by
minimizing the effects of reddening and metallicity on the distance result. Combining our new near-infrared period-luminosity relations 
with the previous optical photometry we obtain a true distance modulus to NGC 7793 of $(27.66 \pm 0.04)$ mag (statistical) $\pm 0.07$ mag (systematic), i.e. a distance 
of $(3.40 \pm 0.17)$ Mpc. We also determine the mean reddening affecting the Cepheids 
to be $E(B-V)=(0.08 \pm 0.02)$ mag, demonstrating that there is significant dust extinction 
intrinsic to the galaxy in addition to the small foreground extinction. A comparison of the new, improved Cepheid distance to earlier 
distance determinations of NGC 7793 from the Tully-Fisher and TRGB methods yields agreement within the reported uncertainties of these previous
measurements.
\end{abstract}


\section{Introduction}

The Araucaria Project is an international project focusing on precise calibration of the  cosmic distance scale, using a variety of stellar distance indicators (Gieren et al. 2005a). The calibration
of the first rungs of the distance ladder still resembles the most important problem in the determination of
the Hubble constant (e.g. Freedman et al. 2001; Riess et al. 2016).
Nearby galaxies like those in the Local Group or the Sculptor Group provide perfect environments for improving 
stellar methods of distance measurement and determine their dependence on metallicity. Important results have been
achieved by our project in the recent past; for instance, we measured the distance to the Magellanic Clouds
with an unprecedented accuracy using eclipsing binaries, 2\% in the case of the LMC (Pietrzynski et al. 2013), and 3\%
in the case of the SMC (Graczyk et al. 2014). With these results, the Magellanic Clouds, and in particular the
LMC, are now firm anchors for the extragalactic distance scale. Our project has also shown that combined optical and
near-infrared photometry of classical Cepheid variables provides an excellent way to measure the distances to
nearby spiral or irregular galaxies with a typical precision of 3\% (e.g. NGC 300, Gieren et al. 2005b; NGC 55, Gieren et al. 2008; NGC 247, 
Gieren et al. 2009; IC 1613, Pietrzynski et al. 2006; NGC 3109, Soszynski et al. 2006 and M 33, Gieren et al. 2013). 

In a previous study, we reported on the discovery of a sample of classical Cepheids in the Sculptor Group spiral galaxy
NGC 7793 and derived a first Cepheid distance to this galaxy based on these objects to be $27.68 \pm 0.05$ mag (statistical) 
$\pm 0.08$ mag (systematic) (Pietrzynski et al. 2010; hereafter
Paper I). The obvious drawback of this distance determination from optical (V,I) data was its relatively high sensitivity
to dust extinction, including dust extinction produced in the galaxy itself which must affect the Cepheids' luminosities. 
In addition, evidence has been mounting in recent years that the effect of metallicity on Cepheid absolute magnitudes is smaller
in near-infrared than in optical bands, making a distance measurement from near-infrared photometry very
desirable. We therefore obtained near-infrared follow-up photometry of the Cepheids, as reported in section 2
of this paper. An accurate distance to NGC 7793 from Cepheids is of high importance because this galaxy has previous distance determinations
from the TRGB and Tully-Fisher methods whose accuracy can be checked 
by comparing with the Cepheid distance. Moreover, the Type IIP supernova SN 2008bk has been observed in the galaxy and an
improved Cepheid distance will lead to a more accurate determination of its absolute peak magnitude.
Also, the previous Cepheid distance measurement from optical data indicated that NGC 7793 might be the most distant 
spiral galaxy in the Sculptor Group, arguing for a significant extension of the Group in the line of sight, a conclusion
that we wanted to confirm with an improved Cepheid distance from infrared data.

\section{Observations, Data Reduction and Calibration}

We collected our $J$- and $K$-band data under photometric conditions using the High Acuity Wide field K-band Imager (HAWK-I) installed on the 8.2m ESO 
Very Large Telescope at Paranal as a part of a program 087.D-0425(B). The field of view of HAWK-I is 7.5 x 7.5 arcmin
and the pixel scale is 0.106 arcsec/pix. The detector of HAWK-I consists of four chips with a 15 arcsec 
cross-shaped gap between them. We observed one field during two photometric nights on September 5 
and September 18, 2011. The location of this field is shown in Figure 1. It was chosen such as to cover as many 
of the previously 
known Cepheids as possible (13 out of 17). The periods of the Cepheids and a preliminary distance from the optical data 
were published in Paper I. 

From the 13 Cepheids in NGC 7793 located in our HAWK-I field, we were able to identify eleven objects. One Cepheid
fell into a gap between the different chips of the HAWK-I imager, and another Cepheid, located closer to the center
of the galaxy than any other of the variables, could not be clearly identified because
the field in that region was too crowded in the near-infrared.

The HAWK-I observations in the $J$- and $K$- bands were performed using a dithering technique with a 
jitter radius of 30 - 40 arcsec in order to subtract  the  sky. For each band, 30 exposures were taken resulting 
in 30 minutes of total net exposure. To avoid saturation, each exposure performed in the J (K) filter consisted 
of six (four) sub-integrations of 10 seconds (15 seconds) each.

In order to reduce our data, we used the dedicated ESO HAWK-I pipeline implemented in Esorex software and 
corresponding optional Gasgano graphic interface. It includes all necessary procedures such as basic reduction, 
sky subtraction, correction for distortion of the detector and stacking into the final images. 
We performed point-spread function (PSF) photometry of our images using DAOPHOT and ALLSTAR, as described
in Pietrzynski et al.(2002) after dividing each multi-extension fit into four ordinary fits corresponding to each chip 
of the detector. We carefully selected about  7-20 isolated stars to calculate the PSF models on each chip.

The calibration of the photometry onto the standard system was based on the
observations of 14 standard stars from the United Kingdom Infrared Telescope ( UKIRT )
list (Hawarden et al. 2001) observed on 2 photometric nights. All of them were observed
along with the target fields in NGC 7793 at different air masses and under photometric
conditions. We obtained the absolute photometric zero points with an accuracy close to
0.03 mag, in both filters. This is demonstrated in Figure 2 which shows the histograms of the measured magnitude differences for constant stars in our fields 
between the two nights.

In addition we analysed IR images of NGC 7793 which we obtained with the ISAAC near-infrared camera on the ESO VLT
as part of ESO proposal 171.D-0004(A).
These observations were obtained under photometric conditions but the integration times for both J and K bands 
were  too short to obtain accurate photometry of the Cepheids in this galaxy. However, they are very well suited for checking 
the zero points of our HAWK-I photometry. In order to do that we reduced the ISAAC data in the same manner as reported in our 
previous papers (e.g. Gieren et al. 2005b). The calibration to the UKIRT system was based on 7 standard stars. 
The photometric zero points were determined with an accuracy better than 0.03 mag for both J and K filters. 

The observed ISAAC field overlaps with all HAWK-I chips. We cross-matched common stars in both photometric lists and calculated the mean 
difference between ISAAC and HAWKI photometry. In all cases the differences in the zero points in both J and K filters, and for all 
HAWK-I chips were smaller than 0.03 mag, a result which strongly reinforces both calibrations.

\section{Results}

\subsection{Near-infrared photometry and mean magnitudes of the Cepheids} 
In the resulting near-infrared photometric catalog we were able to identify, as stated above, 11 Cepheids previously
discovered by Pietrzynski et al. (2010). These variables span a period range from 26 to 62 days. Figure 3 shows their
location in the $K$ versus $J-K$ color-magnitude diagram obtained from our data. 
As one can appreciate, the Cepheids fall into the expected region of the HRD and do fill the 
instability strip rather homogeneously, showing that no significant selection bias affects our period-luminosity relations. It is also worthwhile noticing that the faintest Cepheid in our sample is still about 3 mag brighter than the faintest stars measured in the field. Therefore we do not expect that our near-infrared period-luminosity relations are significantly affected by a Malmquist bias.

In order to calculate the Cepheid mean magnitudes, we simply took the average of the two observations per Cepheid
obtained in this study (except the object cep012 which
has only one measurement in J and K each, which we adopted as its respective mean
magnitudes). Table 1 presents  the detailed  journal of individual observations of our 11 Cepheids.
As can be seen, the four Cepheids with short periods (26-28 days) and two observations are brighter on September 18 
than on September 5. 
It should be noted that for these variables the phase progression between the first and second observing night
was just about half a pulsation cycle, which can lead to a large difference in their magnitudes on these nights,
a fact we have confirmed by simulations. This shows that even having a reasonably large sample 
of Cepheids it is important to obtain more than 
one observation in order to determine accurate mean magnitudes for the individual variables.

In principle one could follow the procedures described by Soszynski et al. (2005) and Inno et al. (2015) which allow 
to calculate the mean magnitude of a Cepheid in near-infrared bands from a single random-phase NIR magnitude measurement 
if accurate optical light curves and period of the Cepheid are available. The method 
however relies critically on well-known phase shifts between the optical and near-infrared observations. 
In our case, the time 
interval between our previous optical datasets and the current near-infrared follow-up observations of the Cepheids is 
too large (6-8 years)
to apply this technique, so we decided to adopt the average the of the two measured magnitudes to estimate the mean magnitudes.

\subsection{Near-infrared period-luminosity relations and distance determination}
Figure 4 presents the  Cepheid period-luminosity (PL) relations for the $J$- and $K$- band in NGC 7793 obtained
from the data in Table 1. The scatter of the Cepheid magnitudes around the mean relations is of 0.13 mag and 0.16 mag for $J$- and $K$- bands respectively and there is no obvious outlier which would hint at a possible misidentification, or any other problem
with the measurements for our Cepheids.

Fitting least-squares lines with two free parameters to the data we obtained slope values of the PL relations 
of $-3.47 \pm 0.30$ and $-3.17 \pm 0.38$ for $J$ and $K$ respectively. Within their large uncertainties, these values
are in agreement with the corresponding  slopes of the near-infrared Cepheid PL relations in the LMC obtained by  
Persson et al. (2004) (e.g. $-3.153$ in $J$ and $-3.261$ in $K$ ). We adopt the very accurate Persson's slopes in 
our analysis as we have done in our previous papers in the Araucaria Project, and fit the PL relations for NGC 7793 once more, now only with one free parameter - the zero point. This yields the following PL relations for NGC 7793:
\begin{equation}
J = -3.153 \log P + (25.534 \pm 0.042) {\rm mag}
\end{equation}
\begin{equation}
K = -3.261 \log P + (25.214 \pm 0.051) {\rm mag}
\end{equation}

In order to calculate the distance moduli of NGC 7793 relative to the LMC, we had to determine the shifts between the 
Cepheid PL relations in NGC 7793 and in the LMC in the J and K bands. To do this, we 
converted our magnitudes from the UKIRT to the Near-Infrared Camera and Multi-Object Spectrometer (NICMOS) system on 
which the zero-points of Persson et al. (2004) were calibrated. According to Hawarden et al. (2001) there are only small and 
constant zero-point differences between the UKIRT and NICMOS systems: 0.034 mag and 0.015 mag for $J$ and $K$, respectively.
 
To obtain the absolute distance moduli to NGC 7793, we adopted the very accurate (2\%) LMC distance modulus of 18.493 mag measured by Pietrzynski et al. (2013) which is in excellent agreement with the value of 18.50 we had adopted in our previous Araucaria work. Thanks to this very accurate distance measurement to the LMC we are also able to improve the NGC 7793 distance moduli from the $V$- and $I$- bands given originally in Paper I. Table 2 shows the resulting distance moduli of NGC 7793 obtained in the different photometric bands, together with their respective uncertainties.

Following our previous work (e.g. Gieren et al. 2005b, Pietrzynski et al. 2006), we used all measured, reddened distance moduli 
in optical and near-infrared bands to determine the true distance modulus of NGC 7793 . Adopting the 
Schlegel et al. (1998) reddening law we fit a straight line to the
relation $(m-M)_0=(m-M)_\lambda-A_\lambda=(m-M)_\lambda - E(B-V)R_\lambda$. The least-squares fit shown in Figure 5
yields for the true distance modulus of NGC 7793, and the total reddening affecting its Cepheids, the following results:

\begin{equation}
(m-M)_0=27.66 \pm 0.04
\end{equation}
corresponding to distance to NGC 7793 of $d=3.40$ Mpc.

\begin{equation}
E(B-V)=0.08 \pm 0.02
\end{equation}
The last line in Table 2 shows the true distance moduli calculated for each of the four bands, adopting the Schlegel
et al. reddening law together with the reddening determined from the slope in Figure 5. The agreement between the 
different values is excellent, supporting the accuracy of both the values for the reddened distance moduli, and the
value of the total reddening affecting the NGC 7793 Cepheids of our sample.

\section{Discussion and conclusions}
There are several factors contributing to the systematic uncertainty of the distance determination to NGC 7793
derived in this paper. These include the accuracy of the zero point of the photometry, the accuracy of the distance
to the LMC which we have adopted as the fiducial galaxy to which the Cepheid PL relation in NGC 7793 is tied, the uncertainty related to a metallicity correction due to a possible metallicity dependence of the
Cepheid PL relation in tandem with a possible systematic metallicity difference between the sample of Cepheids in NGC 7793
selected for this study, and the sample of calibrating Cepheids in the LMC from the OGLE project.

As argued in section 2 of this paper, we estimate the uncertainty of our photometric calibration and the
resulting zero points accuracy of around 0.03 mag. Regarding the distance to the LMC we have adopted, it was shown
by Pietrzynski et al (2013) that its total uncertainty, including any systematics, 
is 2.2\%, or 0.05 mag. While the effect of metallicity on Cepheid absolute magnitudes is still under dispute,
evidence has been mounting that the effect is very small, and possible vanishing at near-infrared wavelengths
(Freedman and Madore 2011; Storm et al. 2011; Romaniello et al. 2008). This conclusion is supported by a recent
comparison of the {\it distance difference} between the two Magellanic Clouds (whose Cepheid populations exhibit
a metallicity difference of about 0.4 dex (Luck et al. 1998)) derived from the near-geometrical, metallicity-independent eclipsing
binary method (Pietrzynski et al. 2013, Graczyk et al. 2014), and the distance difference between SMC and LMC as determined
from the magnitude offsets of the observed Cepheid K-band PL relations in the two galaxies: the two determinations agree 
to within 1\% (Wielgorski et al. 2017). We also note that the metallicities of the young stellar populations 
in NGC 7793 and the LMC seem to agree to within 0.1 dex (Van Dyk et al. 2012 and Stanghellini et al. 2015), minimizing
a metallicity correction to our distance result even if the metallicity effect in the J and K bands were much
larger than expected according to the studies cited above. 

Another factor which might affect the Cepheid magnitudes is the effect of crowding and blending in our images. In an
earlier study of our project, we made an exhaustive test of how blending affects the distance to another spiral member
of the Sculptor Group, NGC 300, and concluded that its distance modulus was affected by less than 0.04 mag
(Bresolin et al. 2005). While NGC 7793 is more distant than NGC 300, the HAWK-I imager used in the present study has a higher resolution 
(seeing for our HAWK-I data is about 0.5 arcsec) than the ISAAC near-infrared camera we used in the NGC 300 NIR work (Gieren et al. 2005b), leading to the expectation that
the blending effect on the current NGC 7793 Cepheids should be of a similar size as the one we measured in NGC 300, given also the fact 
that both NGC 300 and NGC 7793 have very similar inclinations (nearly face-on) with respect to the line-of-sight. We adopt 0.04 mag
as a contribution of the effect of crowding and blending to the systematic uncertainty of our current distance determination
to NGC 7793. The total systematic uncertainty on our current distance measurement of NGC 7793 is therefore estimated
to be about 0.07 mag, or 3\%.

Apart from using Cepheids in this work, and previously in Paper I for a distance determination to NGC 7793, the distance 
to NGC 7793 has also been determined
by applying the TRGB method (Karachentsev et al. 2003; Jacobs et al. 2009) and the Tully-Fisher method (Tully et al 2009). Jacobs et al. derived 
a distance of $27.79 \pm 0.08$ mag. This result is consistent with ours within a precision of $2 \sigma$.
Another TRGB distance modulus of (27.96 $\pm$ 0.24 mag) was obtained by Karachentsev et al. Within its relatively large uncertainty it also agrees 
with our measurement. The Tully-Fisher determination of the distance to NGC 7793 which yields a distance modulus of $28.06 \pm 0.35$ mag (Tully et al. 2009), 
is consistent with our current Cepheid determination but has a much larger uncertainty.

Our measurement confirms the large difference between the distances to the Sculptor Group galaxies NGC 300 - $(m-M)_0=26.37 \pm 0.04 \pm 0.03$ mag 
or $d=1.88 \pm 0.06$ Mpc (Gieren et al. 2005b), NGC 55 - $(m-M)_0=26.434 \pm 0.037$ mag or $d=1.88 \pm 0.03$ Mpc 
(Gieren et al. 2008) and NGC 7793, confirming the elongated structure of the Sculptor Group in the line of sight. It might even
be possible that NGC 7793 does not belong to the Group, but a discussion of this possibility is beyond the scope of this study.

As the principal result of this study, we confirm, and significantly improve the accuracy of the distance determination to NGC 7793 obtained in Paper I.
Extending our wavelength baseline to 2.2 micron we confirm that there is significant intrinsic reddening in NGC 7793 which dominates
the total reddening affecting the Cepheids in the galaxy, and which will need to be taken into account in any other photometric
distance determination to this galaxy. The current distance measurement to NGC 7793 from optical and near-infrared photometry
of Cepheid variables is estimated to be accurate to about 5\%.

\acknowledgments
We would like to thank the anonymous referee for constructive and helpful comments. 

The research leading to these results has received funding from the European Research Council (ERC) under 
the European Union's Horizon 2020 research and innovation program (grant agreement No 695099).
WG, MG and DG gratefully acknowledge financial support for this work from  the Millenium Institute of Astrophysics (MAS) 
of the Iniciativa Cientifica Milenio del Ministerio de Economia, Fomento y Turismo de Chile, project IC120009.
We (WG, GP and DG) also very gratefully acknowledge financial support for this work from the BASAL Centro de Astrofisica
y Tecnologias Afines (CATA) PFB-06/2007. We also acknowledge support from 
the IdP II 2015 0002 64 grant of the Polish Ministry of Science and Higher Education.  
Last not least, we are grateful to the staff at Paranal Observatory (ESO) for their excellent support.

%\appendix

%\section{Appendix material}

\clearpage

\begin{figure}
\plotone{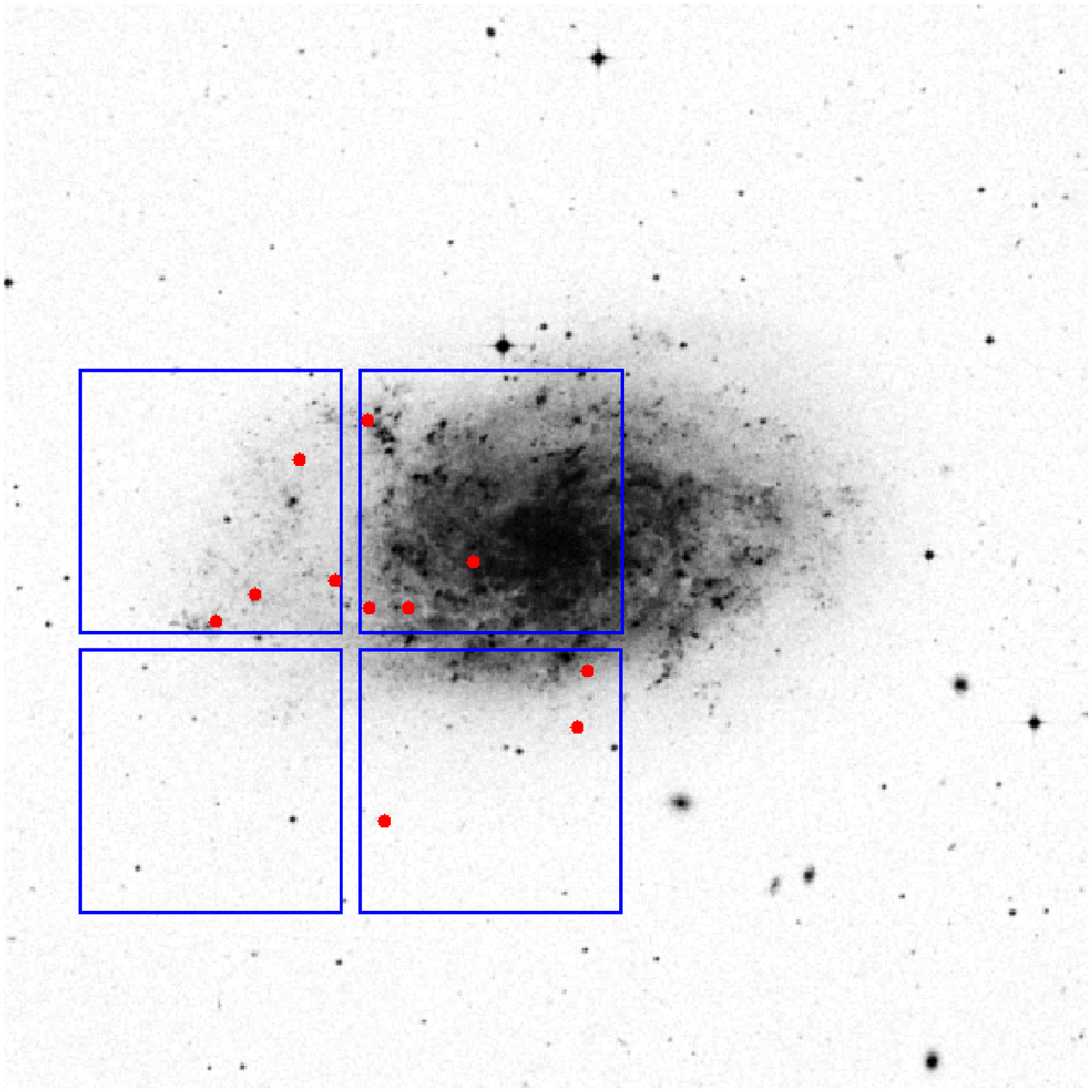}
\caption{Location of the observed HAWK-I field in NGC 7793 which is depicted by the blue squares (four chips with a total field
of view of 7.5x7.5 arcmin). Red dots mark the positions of the eleven Cepheids used in this study.  The field of view of the image is 15x15 arcmin. North is up, and East is left.\label{fig1}}
\end{figure}

\clearpage

\begin{figure}
\plotone{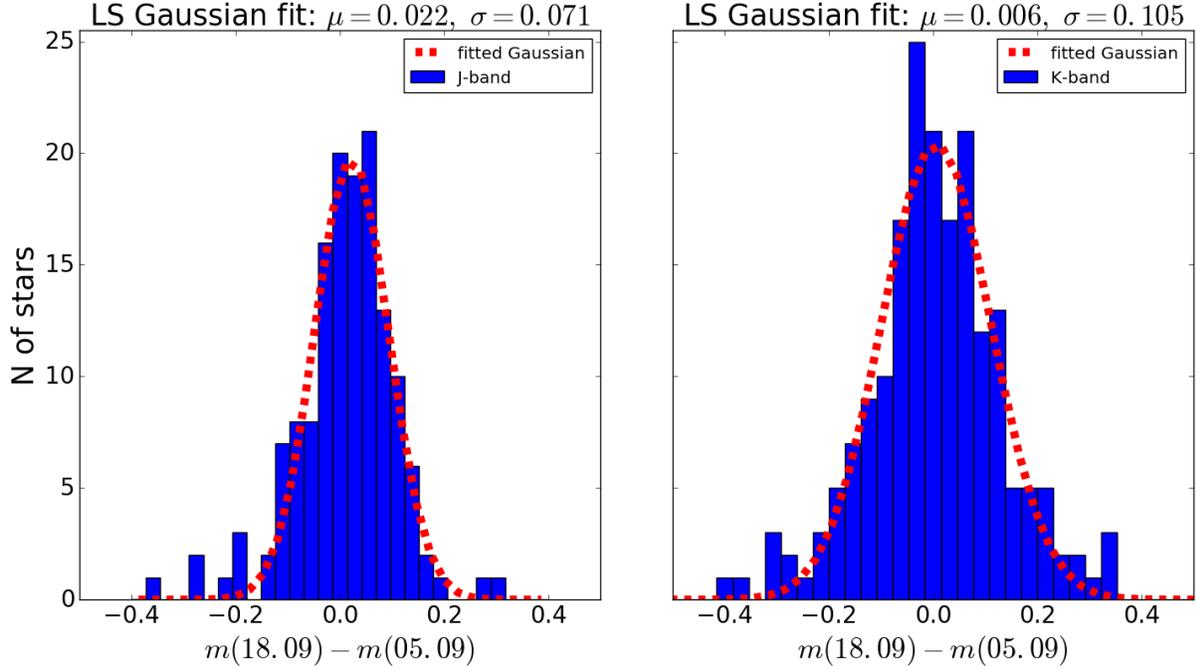}
\caption{Histograms showing the magnitude differences of stars in the field of view of CHIP2 of 
the HAWK-I camera between the two observing nights, for both 
near-infrared filters. Red dashed lines depict best Gaussian fits to the histograms. As can be appreciated, 
the mean values  of the  fitted distributions (representing shifts between zero points of photometry between the two nights) 
are smaller than 0.03 mag. The corresponding plots for the other chips of the camera we used look very similar. \label{fig2}}
\end{figure}

\clearpage

\begin{figure}
\includegraphics{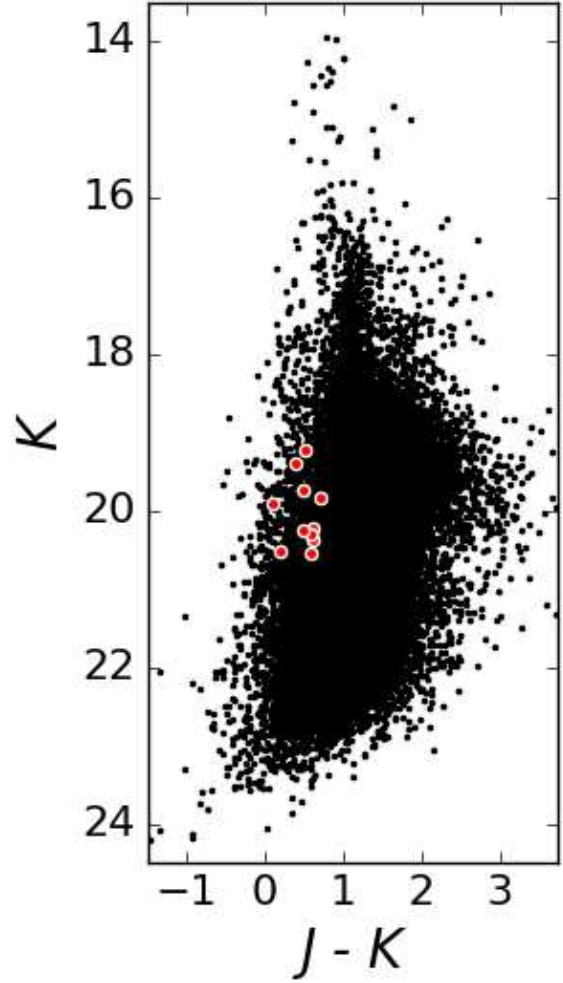}
\centering
\caption{Near-infrared color-magnitude diagram of NGC 7793 stars with photometry from our HAWK-I data, showing the location of the Cepheids
with red dots. 
The variables delineate the classical Cepheid instability strip. The faintest stars we could measure
are about 3 mag less luminous than the faintest Cepheid in our sample.}
\end{figure}

\clearpage

\begin{figure}
\includegraphics[scale=.8]{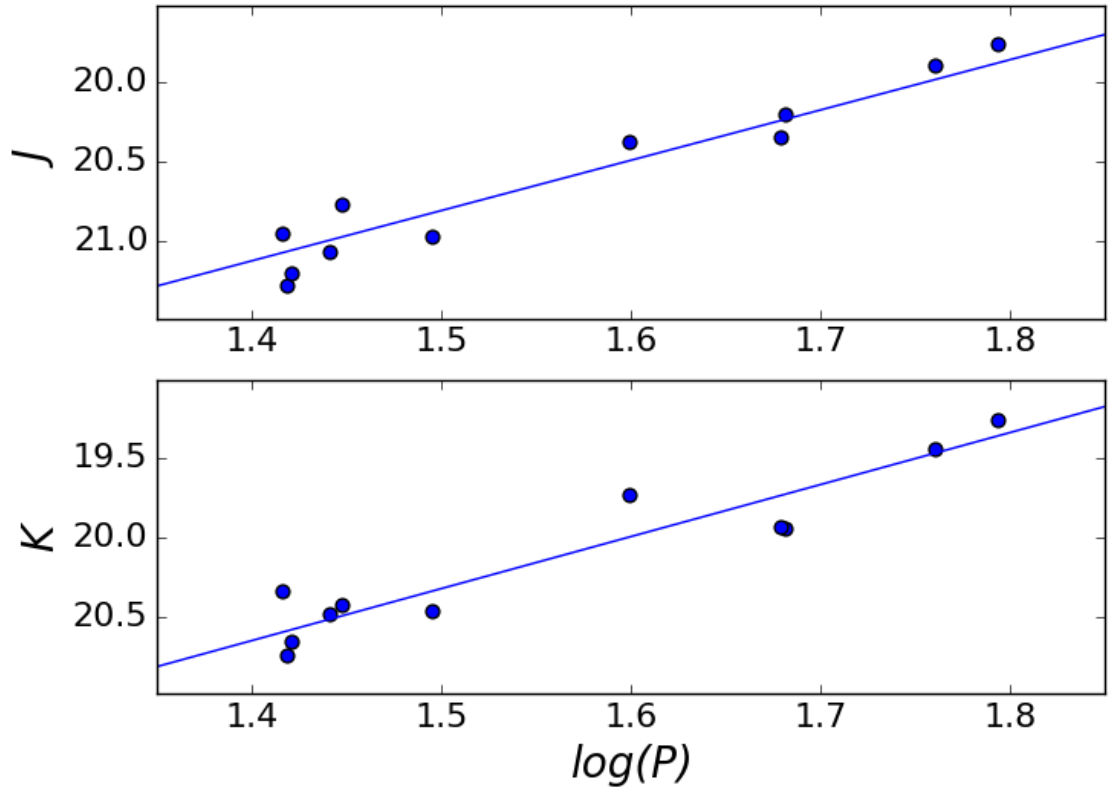}
\centering
\caption{The near-infrared period-luminosity relations in the $J$ and $K$ bands as defined by the Cepheids in NGC 7793. 
The mean magnitudes of each variable were obtained by averaging two random-phase observations.
The respective slopes on these diagrams were adopted from the LMC Cepheids (Persson et al. 2004).}
\end{figure}

\clearpage

\begin{figure}
\includegraphics[scale=.85]{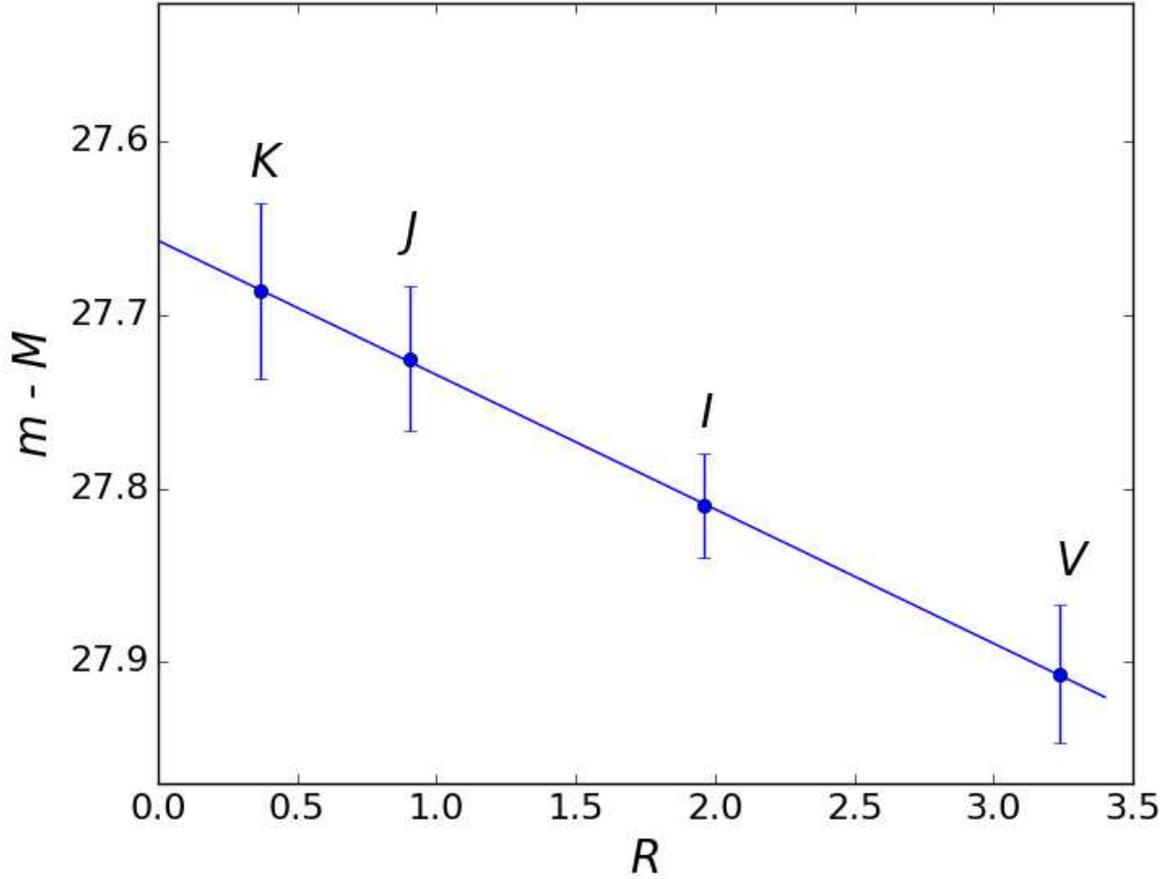}
\centering
\caption{Apparent distance moduli of NGC 7793 as derived in different photometric bands, plotted against the ratio of total
to selective extinction as adopted from the Schlegel et al. (1998) reddening law. The intersection and slope of the best-fitting line
give the true distance modulus and reddening, respectively. The extremely small scatter of the reddened distance moduli
in the different bands around the best-fitting line is spurious.}
\end{figure}

\begin{deluxetable}{lccccccc}
\tablewidth{0pt}
\tablecaption{Journal of $J$- and $K$- band individual observations of NGC 7793 Cepheids. Names of objects and periods (accurate to $10^{-3} \times$ P)
were adopted from Pietrzynski et al. (2010). }
\tablehead{
\colhead{ID} & \colhead{P}&
\colhead{HJD-2450000} & \colhead{$J$}  &
\colhead{$\sigma_J$} & \colhead{HJD-2450000}  &
\colhead{$K$}  & \colhead{$\sigma_K$}  \\
\colhead{} & \colhead{(days)} & \colhead{($J$ observations)} &
\colhead{(mag)} & \colhead{(mag)} & \colhead{($K$ observations)} &
\colhead{(mag)} & \colhead{(mag)}
}
\startdata
cep002 & 62.120 & 5809.68 & 19.758 & 0.014 & 5809.65 & 19.292 & 0.032 \\
&  & 5822.71 & 19.753 & 0.013 & 5822.67 & 19.232 & 0.030 \\
cep003 & 57.600 & 5809.68 & 19.997 & 0.012 & 5809.65 & 19.497 & 0.023 \\
& & 5822.71 & 19.797 & 0.009 & 5822.67 & 19.402 & 0.024 \\
cep006 & 48.008 & 5809.68 & 20.415 & 0.022 & 5809.65 & 19.996 & 0.050 \\
& & 5822.71 & 19.995 & 0.022 & 5822.67 & 19.898 &  0.039 \\
cep007 & 47.783 & 5809.68 & 20.462 & 0.012 & 5809.65 & 20.134 & 0.031 \\
& & 5822.71 & 20.232 & 0.010 & 5822.67 & 19.738 & 0.023 \\
cep008 & 39.741 & 5809.68 & 20.190 & 0.033 & 5809.65 & 19.633 & 0.049 \\
&  & 5822.71 & 20.556 & 0.037 & 5822.67 & 19.834 & 0.055 \\ 
cep010 & 31.255 & 5809.68 & 21.185 & 0.017 & 5809.65 & 20.668 & 0.044 \\
& & 5822.71 & 20.753 & 0.012 & 5822.67 & 20.256 & 0.036 \\
cep011 & 28.011 & 5809.68 & 20.834 & 0.031 & 5809.65 & 20.341 & 0.065 \\
& & 5822.71 & 20.700 & 0.030 & 5822.67 & 20.508 & 0.078 \\
cep012 & 27.601 & 5809.68 & 21.062 & 0.026 & 5809.65 & 20.483 & 0.058  \\
cep013 & 26.364 & 5809.68 & 21.413 & 0.023 & 5809.65 & 20.931 & 0.063 \\
&  & 5822.71 & 20.982 & 0.018 & 5822.67 & 20.374 & 0.048 \\
cep014 & 26.216 & 5809.68 & 21.409 & 0.021 & 5809.65 & 20.926 & 0.055 \\
& & 5822.71 & 21.133 & 0.017 & 5822.67 & 20.546 & 0.052 \\
cep015 & 26.083 & 5809.68 & 21.062 & 0.035 & 5809.65 & 20.445 & 0.060 \\
& & 5822.71 & 20.836 & 0.029 & 5822.67 & 20.228 & 0.053
\enddata

\end{deluxetable}

\begin{deluxetable}{cccccc}
\tablewidth{0pt}
\tablecaption{Distance moduli of NGC 7793 from different bands, with ratios of total to selective extinction adopted from Schlegel et al. (1998). All quantities in mag.}
\tablehead{
\colhead{Band} & \colhead{$V$} & \colhead{$I$} & \colhead{$J$} & \colhead{$K$} & \colhead{$E(B-V)$}}
\startdata
(m-M) & 27.907 $\pm $ 0.040 & 27.810 $\pm$ 0.030 & 27.725 $\pm$ 0.042 & 27.686 $\pm$ 0.051 & ...\\
R & 3.240 & 1.962 & 0.902 & 0.367 & ...\\
(m-M)$_0$ & 27.65 & 27.65 & 27.65 & 27.66 & 0.08 \\
\enddata
\end{deluxetable}

\end{document}